\documentclass[12pt]{article}
\usepackage{graphicx}

\newcommand\pubdate{\today}

\newcommand{\Bx}{x_{\rm B}}

\def\biselli{Department of Physics\\
Fairfield University, Fairfield CT, 06611}

\def\Title#1{\begin{center} {\Large #1 } \end{center}}
\def\Author#1{\begin{center}{ \sc #1} \end{center}}
\def\Address#1{\begin{center}{ \it #1} \end{center}}

\newcommand\pubblock{\rightline{\begin{tabular}{l} 
         \pubdate  \end{tabular}}}
\newenvironment{Abstract}{\begin{quotation}  }{\end{quotation}}
\newenvironment{Presented}{\begin{quotation} \begin{center} 
             PRESENTED AT\end{center}\bigskip 
      \begin{center}\begin{large}}{\end{large}\end{center} \end{quotation}}

\def\beq{\begin{equation}}
\def\eeq#1{\label{#1}\end{equation}}
\def\eeqn{\end{equation}}

\def\beqa{\begin{eqnarray}}
\def\eeqa#1{\label{#1}\end{eqnarray}}
\def\eeqan{\end{eqnarray}}

\let\bar=\overbar

\def\Dslash{\not{\hbox{\kern-4pt $D$}}}
\def\dslash{\not{\hbox{\kern-2pt $\del$}}}

\def\msb{{\bar{\ssstyle M \kern -1pt S}}}

\begin{document}
\begin{titlepage}
\pubblock

\vfill
\Title{Deeply virtual Compton scattering at Jefferson Laboratory}
\vfill
\Author{ Angela Biselli\\for the CLAS collaboration}
\Address{\biselli}
\vfill
\begin{Abstract}
The generalized parton distributions (GPDs) have emerged as a universal tool to describe hadrons in terms of their elementary constituents, the quarks and the gluons. Deeply virtual Compton scattering  (DVCS) on a proton or neutron ($N$), $e N \rightarrow e' N' \gamma$, is the process more directly interpretable in terms of GPDs. The amplitudes of DVCS and Bethe-Heitler,  the process where a photon is emitted by either the incident  or scattered electron,  can be accessed via cross-section measurements or exploiting their interference which gives rise to spin asymmetries. Spin asymmetries, cross sections and cross-section differences can  be connected to different combinations  of the four leading-twist GPDs (${H}$, ${E}$, ${\tilde{H}}$, ${\tilde{E}}$) for each quark flavors, depending on the observable and on the type of target.  
 
This paper gives an overview of recent experimental results obtained for DVCS at Jefferson Laboratory  in the halls A and  B. Several experiments have been done extracting DVCS observables over large kinematics regions. Multiple measurements with overlapping kinematic regions allow to perform a quasi-model independent extraction of the Compton form factors, which are GPDs integrals, revealing a 3D image of the nucleon. 


\end{Abstract}
\vfill
\begin{Presented}
CIPANP 2015\\
Vail CO,  May 19--24, 2015
\end{Presented}
\vfill
\end{titlepage}
\def\thefootnote{\fnsymbol{footnote}}
\setcounter{footnote}{0}

\section{Introduction}
Elastic scattering and deep inelastic scattering have been for years the methods of choice to study the structure of the nucleon. Elastic scattering gives access to the form factors which are related to  the transverse spatial  distribution of quarks, whereas deep inelastic scattering gives access to parton distributions which are longitudinal momentum and spin distribution of quarks. While both these quantities are important, it is clear that they are a subset of more fundamental quantities which encompass all the dimensions in space and momentum. The generalized parton distributions (GPDs) give fully correlated quark distributions in both coordinate and momentum space.
These distributions allow access to crucial information such as the  angular momentum distribution of quarks in the nucleon~\cite{Mueller:1998fv,Ji:1996ek,Radyushkin:1997ki}.

\begin{figure}[htb]
\centering
\includegraphics[height=1.5in]{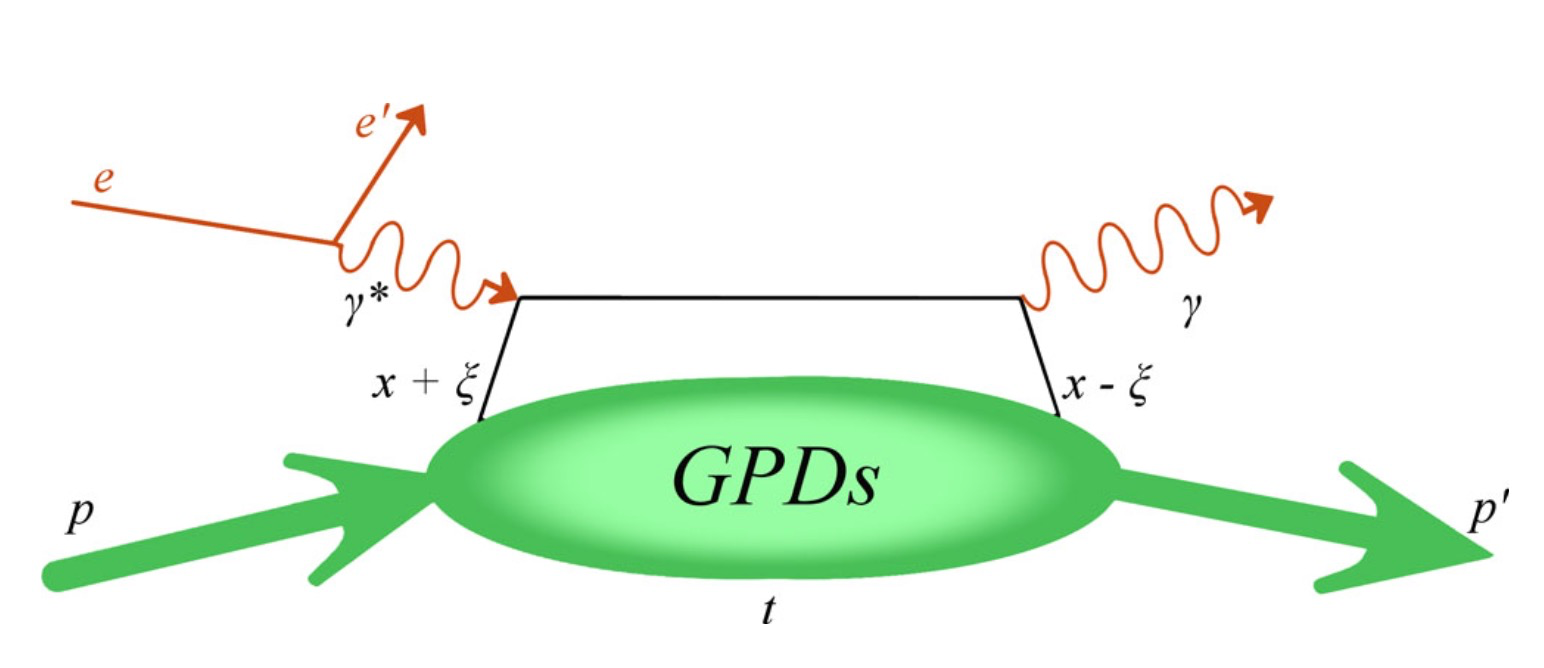}
\caption{Handbag diagram for the DVCS process.}
\label{fig:handbag}
\end{figure}

The cleanest way to access GPDs is via deeply virtual Compton scattering (DVCS), where the virtual photon interacts with a single quark of the nucleon radiating a real photon. As shown in Fig.~\ref{fig:handbag}, this exclusive process can be factorized, at high photon virtualities,  into a hard scattering part, that can be treated perturbatively, and a nucleon-structure part, parameterized by the GPDs.

At  leading twist the soft process is described by four chiral-even GPDs: $H$, $\tilde{H}$, $E$, and $\tilde{E}$, which depend on the longitudinal momentum fraction transferred to the proton, 
$\xi \simeq x_B/(2-x_B)$, the momentum transfer, $t$, between the virtual and the real photons and  
 the momentum fraction  of the struck quark, $x+\xi$, which is not experimentally accessible.  
 All these four GPDs all involve processes that conserve the quark helicity, but while $H$ and $\tilde{H}$ preserve the nucleon helicity, $E$ and $\tilde{E}$ do not. The GPD $H$ ($E$) is an average over the two possible possible quark-helicity conserving processes, whereas $\tilde{H}$ ($\tilde{E}$) is a difference. GPDs are defined at the quark level for each  flavor.

From the experimental point of view,  the DVCS cannot be disentangled  from the Bethe Heitler process, where the final-state photon is emitted by either the incoming or the outgoing electron. To extract the DVCS amplitude $T_{\rm DVCS}$, which contains the GPDs, one can measure both cross-section
\begin{equation}
\frac{d^4\sigma}{dQ^2 dx_B dt d\phi} \propto |T_{\rm DVCS}+T_{\rm BH}|^2=|T_{\rm DVCS}|^{2}+|T_{\rm BH}|^{2}+I,
\end{equation}
 or asymmetries, which at leading twist can be written as:
\begin{equation}
A=\frac{\sigma^+-\sigma^-}{\sigma^++\sigma^-} \propto \frac{I}{|T_{\rm DVCS}|^2+|T_{\rm BH}|^2+I}.
\end{equation} 
 Here  $I= T_{\rm DVCS} T_{\rm BH} ^\ast+ T_{\rm DVCS} ^\ast T_{\rm BH}$ is the interference between the two processes.
 The DVCS amplitude $T_{\rm DVCS}$ depends on linear combinations of Compton form factors $\cal F$, whose real and imaginary parts are connected to the GPDs by
\begin{eqnarray}
\Re{\rm e}{\cal F} & = {\cal P}\int_{-1}^{1}dx\left[\frac{1}{x-\xi}\mp\frac{1}{x+\xi}\right]F(x,\xi,t) \\
\Im{\rm m}\mathcal{F} &= \pi\left[F(\xi,\xi,t)\mp F(-\xi,\xi,t)\right]
\end{eqnarray}
Here the ``$\mp$'' sign  apply, respectively, to the quark-helicity independent, or unpolarized, GPDs $(H,E)$ and to the quark-helicity dependent, or polarized, GPDs $(\tilde{H}, \tilde{E})$.
This means that the experimental observables depend on eight GPD-related quantities. Moreover the CFFs accessed experimentally are not directly the quark ones but CFFs for the type of target, proton, neutron or nuclei, which are linear combination of CFFs of different quark flavors. Luckily different observables (e.g. beam and target asymmetries) and different targets have different sensitivities to the various CFFs and therefore by performing several measurements one can separate the different contributions for a certain target and ultimately, by combining different target measurements one can perform flavor separation.   
For instance, the beam-spin asymmetry  $A_{\rm LU}$ can be expressed as~\cite{Belitsky:2001ns}
\begin{eqnarray}
A_{\rm LU}(\phi) \propto \Im{\rm m}
\left\{F_1 {\cal H} + \frac{\Bx}{2 - \Bx} (F_1 + F_2) (\widetilde {{\cal H}}-\frac{t^2}{4M^2}F_2{\cal E}) +....\right\} \sin \phi \, ,
\end{eqnarray}
where $F_{1}$ and $F_{2}$ are the form factors,
and is sensitive to $\mathcal{H}$, $\tilde{\mathcal{H}}$, and $\mathcal{E}$,  particularly to $\mathcal{H}_{p}$ for the proton and to $\mathcal{E}_{n}$ for the neutron.
The longitudinal target-spin asymmetry $A_{\rm UL}$
\begin{eqnarray}
A_{\rm UL}(\phi)  \propto 
\Im{\rm m}
\left\{
F_1 \widetilde { \cal H}
+
\frac{\Bx}{2 - \Bx} (F_1 + F_2) ({\cal H}+\frac{\Bx}{2}{\cal E}) +....
\right\} \sin\phi \, ,
\end{eqnarray}

is equally sensitive to $\mathcal{H}_{p}$ and $\tilde{\mathcal{H}}_{p}$ for the proton and $\mathcal{H}_{n}$ for the neutron. Furthermore  the cross section and the double spin asymmetry are sensitive to the real part CFFs.


\section{Results}
\begin{figure}[htb]
\centering
\includegraphics[height=3.05in]{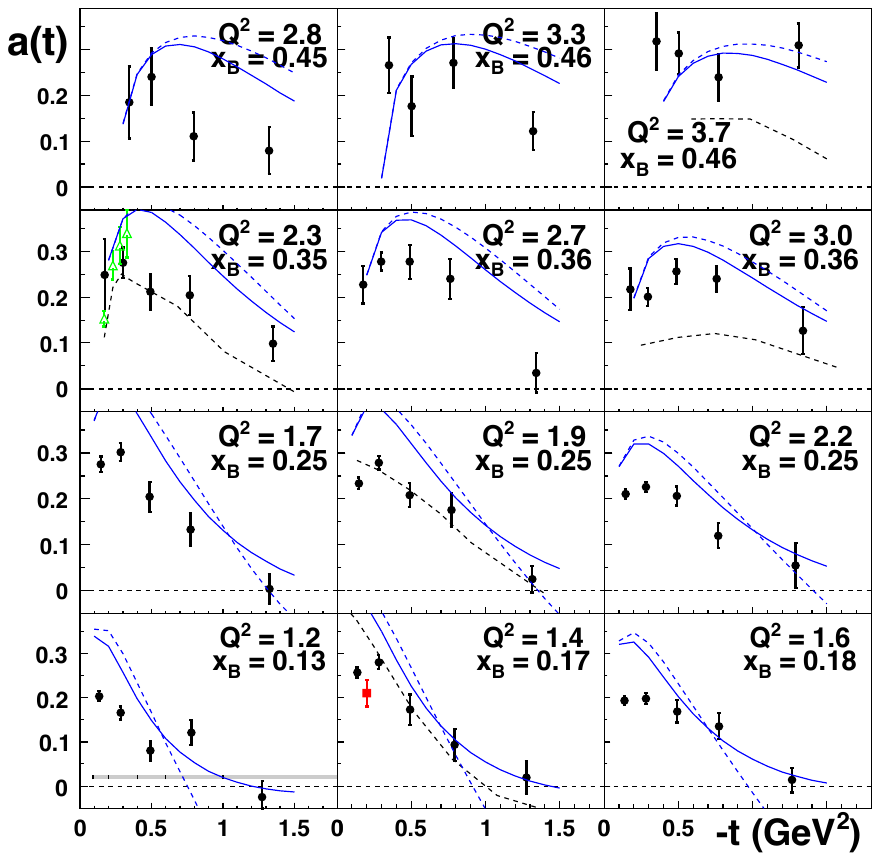}
\includegraphics[height=3.05in]{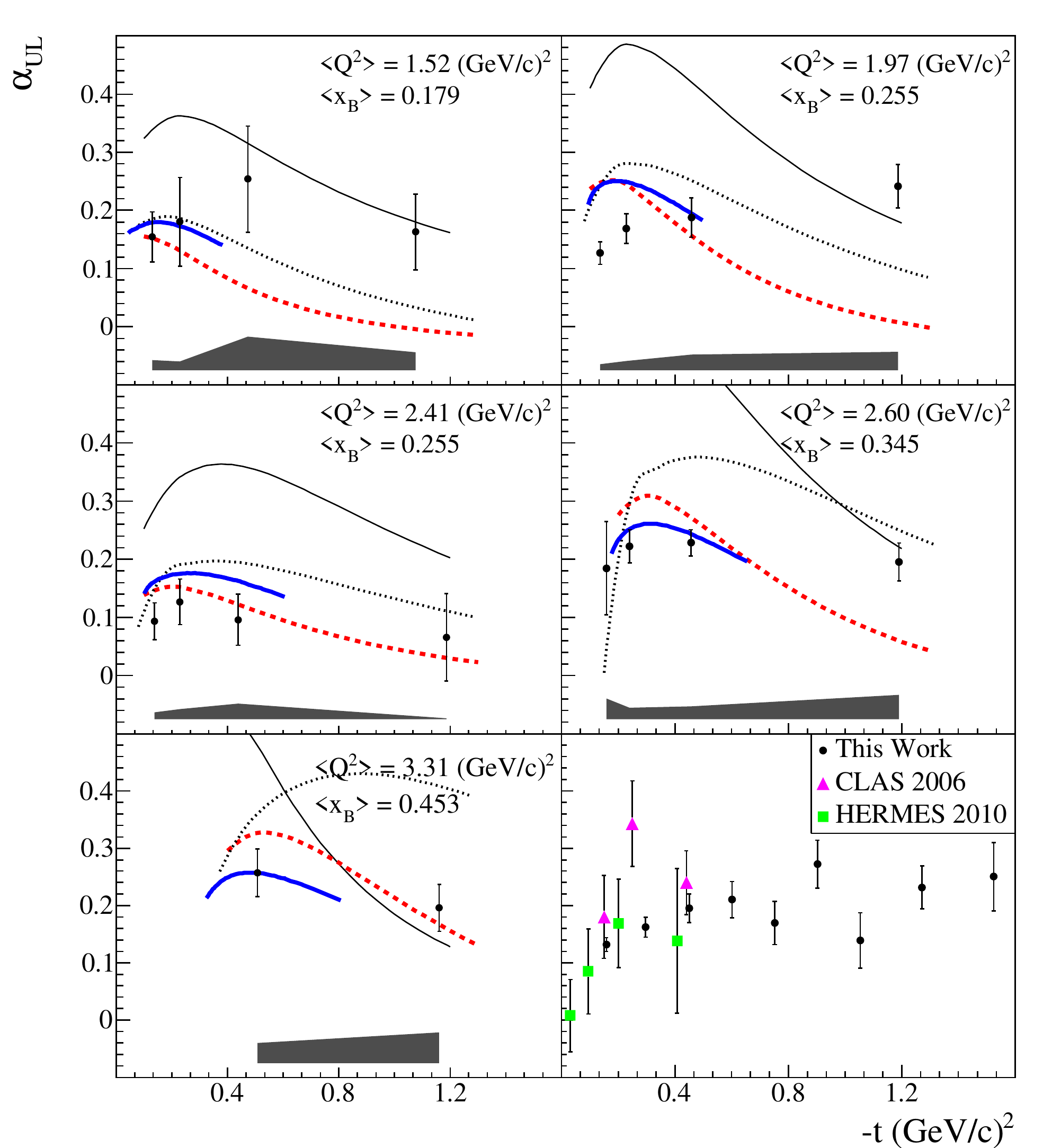}
\caption{BSA (left panel) and TSA (right panel) $-t$ dependence of the $\sin\phi$ term. Left panel: the curves show  a Regge calculation~~\cite{Laget:2007qm} (black dashed), and the VGG prediction at twist-2 (blue solid) and twist-3 (blue dot-dashed)~\cite{Guidal:2004nd}. Right panel: the curves   show the predictions of 4 GPDs  models: VGG~\cite{Guidal:2004nd} (red dashed line), (ii) GK~\cite{Kroll:2012sm} (black dotted lube), KMM12~\cite{Kumericki:2013br} (blue thick solid line), GGL~\cite{Goldstein:2010gu} (black solid line). The data agree qualitatively with the model predictions but clearly provide new constraints to the GPDs. More details on the models comparison can be found in~\cite{Girod:2007jq,Seder:2014cdc}}
\label{fig:BSATSA}
\end{figure}

DVCS measurements have been performed at COMPASS, H1-ZEUS, Hermes and Jefferson Lab. This paper will focus on results obtained at Jefferson Lab where the two experimental halls, A and B, have a unique access to the large $\Bx$ region and therefore have an  insight into the quarks valence region. Several experiments to extract  DVCS asymmetries and/or cross sections on proton, neutron and nuclei targets have been performed over the last decade at Jefferson Lab. Two non-dedicated experiments in Hall B using  the CEBAF large acceptance spectrometer (CLAS)~\cite{Mecking:2003zu}  on unpolarized and polarized proton targets measured  the beam spin asymmetry  (BSA)~\cite{Stepanyan:2001sm} and the target spin asymmetry (TSA) \cite{Chen:2006na}, showing the dominance of the handbag process  and a clear twist-2 $\sin\phi$ dependence. These two experiments were followed by two dedicated experiments  with the addition of the inner calorimeter to detect low-angle photons, focused on the extraction the BSA and TSA over a large kinematic range. Figure~\ref{fig:BSATSA} shows the $-t$-dependence of the $\sin\phi$ term of the BSA \cite{Girod:2007jq} and TSA~\cite{Seder:2014cdc}. The proton BSA,  sensitive to $\Im{\rm m} {\cal H}_{p}$, shows a steeper drop than $\Im{\rm m} {\cal \widetilde{H}}_{p}$. Since the Compton form factors $\Im{\rm m} {\cal H}_{p}$ and $\Im{\rm m} {\cal \widetilde{H}}_{p}$ are related to the Fourier transforms of the electric charge and axial charge respectively, this behavior indicates that the axial charge is more concentrated in the proton center than the electric one. 

\begin{figure}[htb]
\centering
\includegraphics[height=4in]{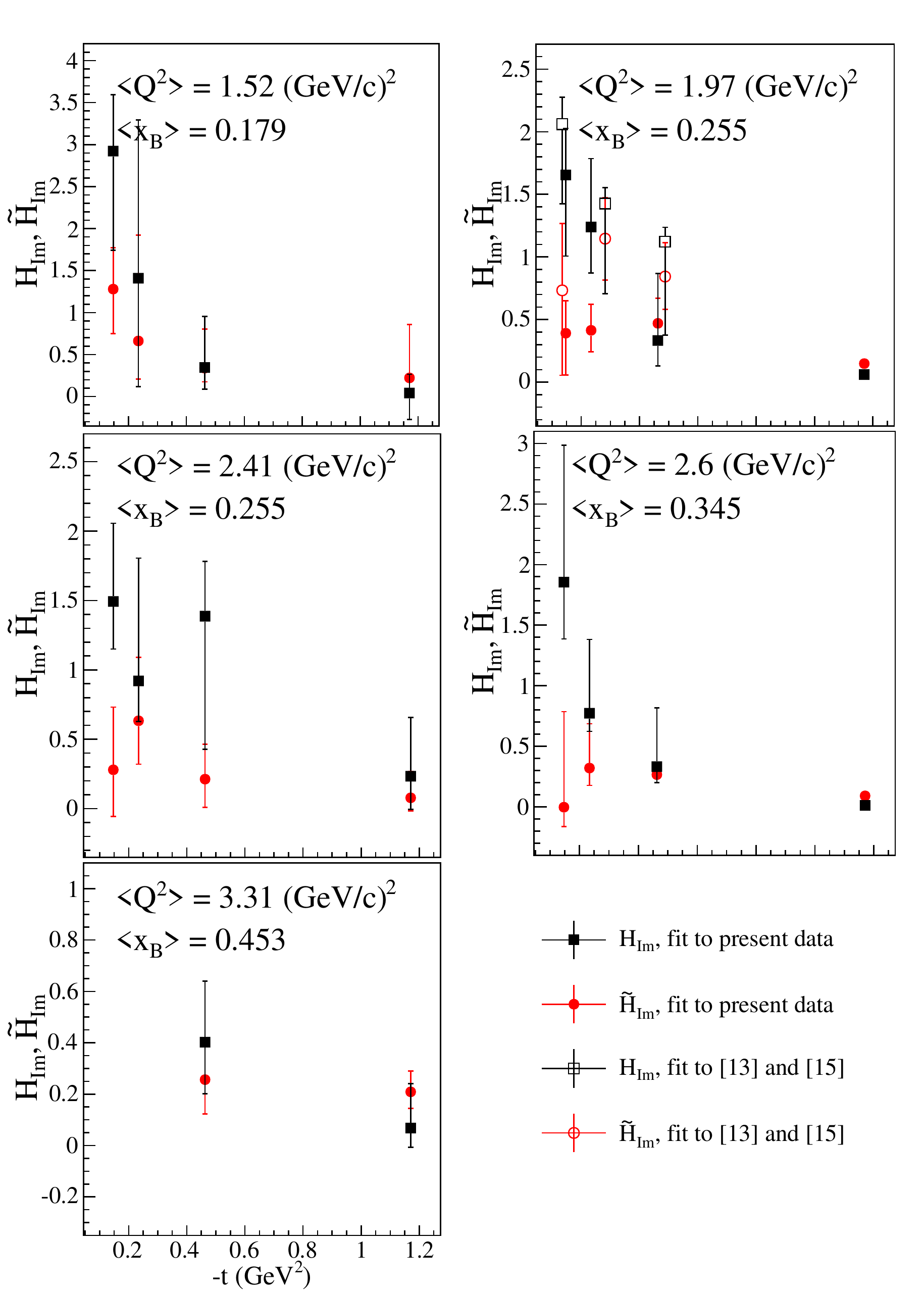}
\caption{Compton form factor extraction from the simultaneous fit of BSA, TSA and DSA. The solid black  squares and the solid red circles show the results for the imaginary parts of ${\cal{H}}_{p}$ and $ \widetilde{\cal{H}}_{p}$ respectively. The results are compared to  fits to previous CLAS data~\cite{Chen:2006na,Girod:2007jq}}
\label{fig:CFF}
\end{figure}

Using the polarized proton data, the BSA and the double spin asymmetry (DSA) were also measured~\cite{Pisano:2015iqa}. The  measurement of the three asymmetries at the same kinematic points allowed a simultaneous fit to extract the Compton form factors for the proton. This was done using a quasi model-independent technique~\cite{Guidal:2008ie} in which the bounds of the  domains of variation of the CFFs is limited to $\pm 5$ times the value  predicted by the VGG model~\cite{Guidal:2004nd}, and $\widetilde{{\cal{ E}}_{p}}$=0. Figure~\ref{fig:CFF} shows the results of the fit for $\Im{\rm m} {\cal H}_{p}$ and $\Im{\rm m} {\cal \widetilde{H}}_{p}$. In addition to confirming the fact that the axial charge is more concentrated than the electrical charge, one can see that the slope of $\Im{\rm m} {\cal H}_{p}$  decreases as $\Bx$ becomes bigger, indicating that the electric charge is more concentrated for valence quarks than sea quarks.

\begin{figure}[htb]
\centering
\includegraphics[height=3.in]{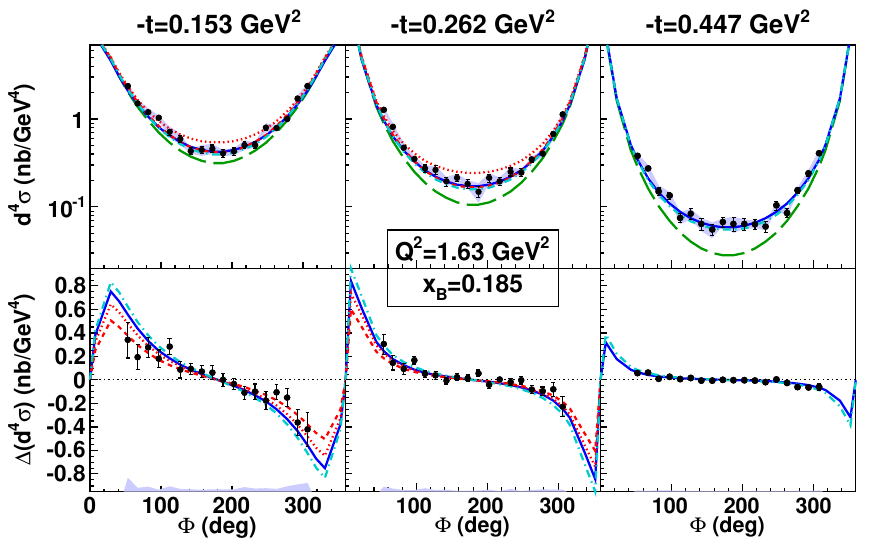}
\caption{Unpolarized cross-section (top row) and beam-polarized cross-section differences  (bottom row) from the CLAS experiment ~\cite{Jo:2015ema} for one kinematic bin $(Q^{2},x_{B})$=(1.63 GeV${}^{2}$,0.185). The three panels in each row correspond to different $-t$ values. The green long-dashed curves correspond to the BH calculation, the other curves corresponds to VGG ~\cite{Guidal:2004nd} (blue solid), KMS ~\cite{Kroll:2012sm} (cyan dash-dotted), and two versions of the KM model ~\cite{Kumericki:2013br} KM10  (red dotted) and KM10a (red short-dashed).}
\label{fig:xsect}
\end{figure}

\begin{figure}[htb]
\centering
\includegraphics[height=3.in]{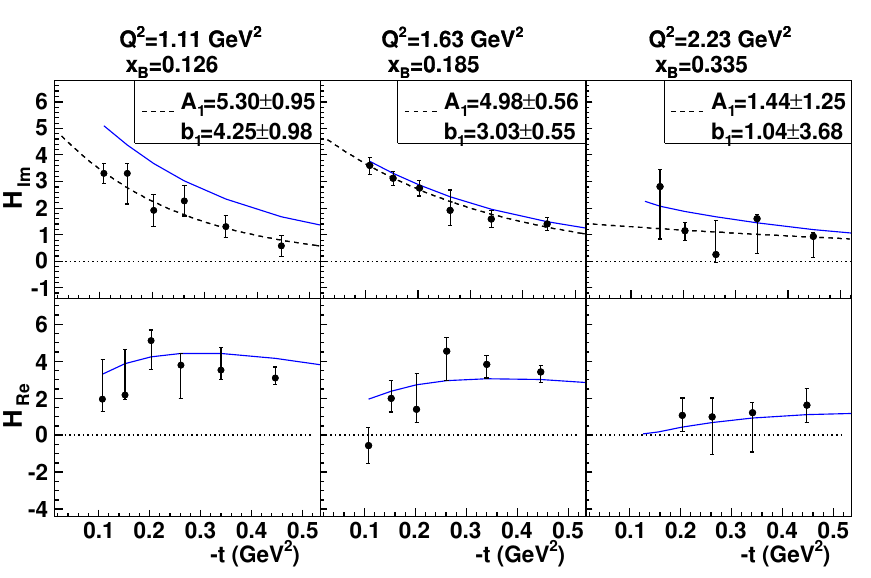}
\caption{Results of the extraction of the Compton form factors $\cal H_{\rm Im}$ (upper panels) and $\cal H_{\rm Re}$ (bottom panels) from the CLAS experiment~\cite{Jo:2015ema} for three of the $(Q^{2},x_{B})$ bins, as a function of $t$. The blue solid curves are the predictions of the VGG model. The black dashed curves show the fit of the results by the function $Ae^{bt}$.
}
\label{fig:xsect2}
\end{figure}

While  beam-spin  and target-spin  asymmetries give access to the imaginary part of the the CFF, double spin asymmetry and cross section measurements are essential to access the real part of the CFF. Both experimental halls at Jefferson Lab have extracted cross sections for the DVCS with proton target. In both experiments  the polarized beam allowed to also measure the beam-polarized cross-section differences. Hall A results from 2006~\cite{MunozCamacho:2006hx} and the recent re-analysis of the same dataset~\cite{Defurne:2015kxq} extracted the cross section and the cross-section differences over a limited $Q^{2}$ and $-t$ with high precision, using a high resolution spectrometer  to detect the electron and the photon with a dedicated calorimeter. A fit of the data allowed to extract both the DVCS and interference Compton form factor, as well as twist-2 and twist-3 contributions. The analysis shows that twist-3 contributions are small and that the cross section shows a large DVCS amplitude. Hall B recently extracted the cross sections and cross-section differences over a large kinematic range~\cite{Jo:2015ema}. A sample of the results for three kinematic bins can be found in Figure~\ref{fig:xsect}, where the comparison with the BH calculation (green curve) shows a the non-zero contribution from the DVCS process. Figure~\ref{fig:xsect2} shows the real and imaginary part of $\cal{H}$ found by fitting the data  using~\cite{Guidal:2008ie} with $\mathcal{E}_{p}$ and $\widetilde{\cal{E}}_{p}$  set to zero. The $\Bx$ trend of the fit indicate that the transverse size and partonic content are bigger at smaller momentum fractions.

\begin{figure}[htb]
\centering
\includegraphics[height=2.0in]{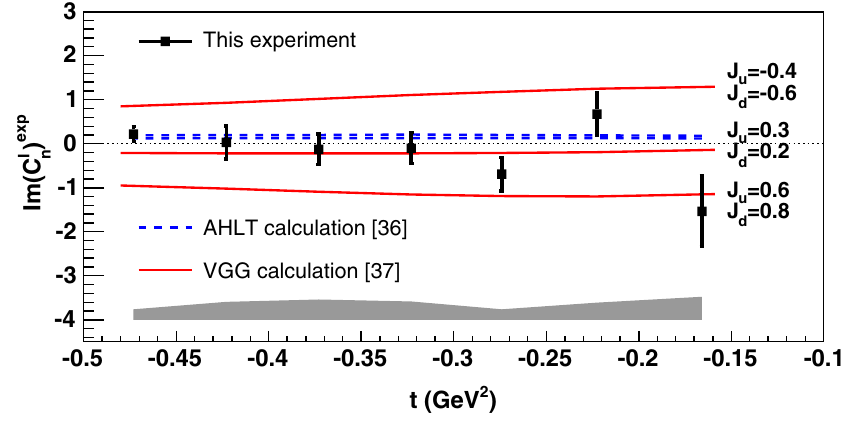}
\caption{$t$ dependence of $\sin\phi$ moment for the neutron. Results are compared to model calculations of neutron GPDs~\cite{Ahmad:2006gn} (blue dashed curve) and ~\cite{Guidal:2004nd} (solid red curves). The three  red solid curves correspond to calculations for three different values of of the $u$ ($J_{u}$ and $d$ ($J_{d}$) quark contributions.}
\label{fig:neutr}
\end{figure}

\begin{figure}[htb]
\centering
\includegraphics[height=1.5in]{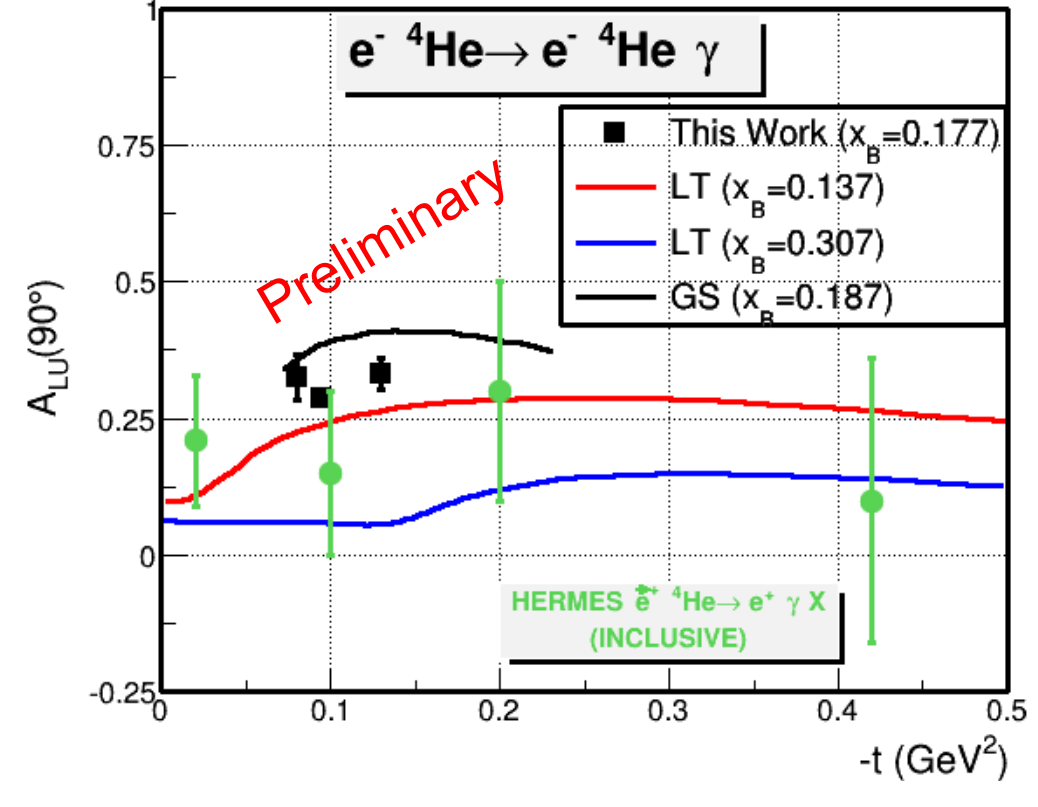}
\includegraphics[height=1.5in]{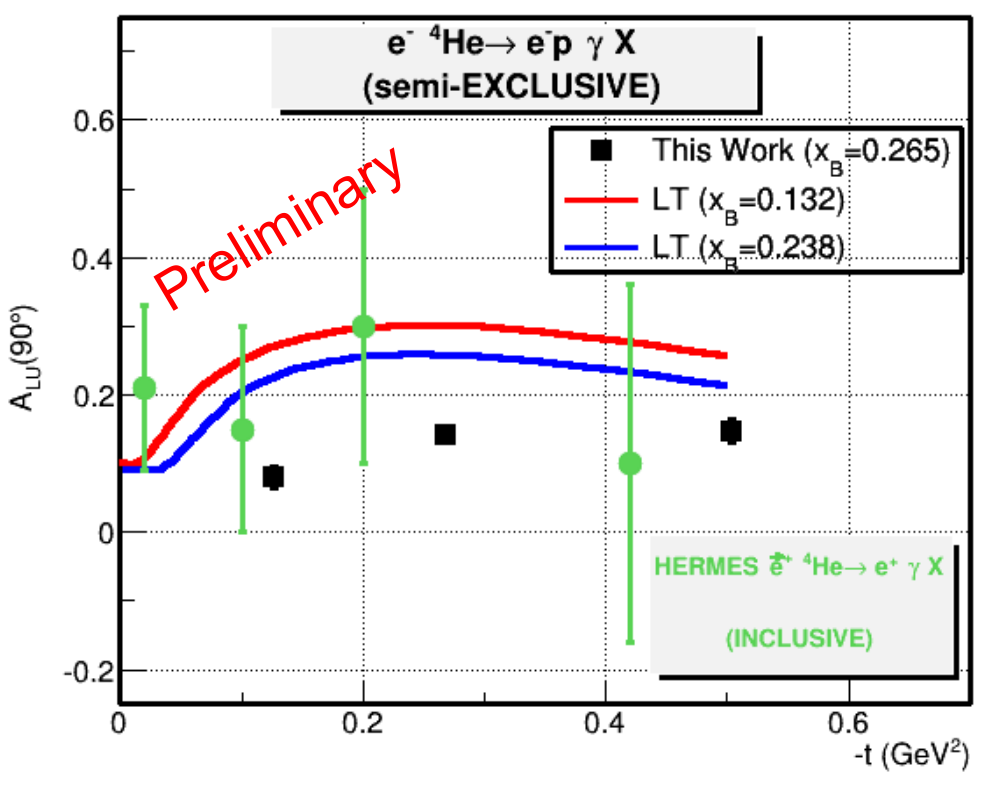}
\caption{Sample of results for DVCS on ${}^{4}{\rm He}$ for coherent DVCS $e {}^{4}{\rm He} \rightarrow e {}^{4}{\rm He} \gamma $ (left panel) and in-coherent  DVCS $e {}^{4}{\rm He} \rightarrow e p \gamma X$ (right panel)}
\label{fig:nuclei}
\end{figure}

The DVCS process on the neutron has been studied as well. Neutron measurements are essential since they are needed, once combined with the proton measurements, to do flavor separation. Hall A has extracted the imaginary part of the DVCS amplitude of the neutron and deuteron by measuring the helicity dependent $D(\vec{e},e',\gamma)N$ cross sections ~\cite{Mazouz:2007ka}. The experiment measured deuteron cross section, which is the sum of coherent deuteron scattering and incoherent neutron and proton scattering, and used liquid ${\rm H}_{2}$ data taken at the same kinematics to subtract the proton quasielastic contribution. The resulting contribution from deuteron and neutron was found compatible with zero.
By fitting the data, the deuteron and neutron $\sin\phi$ moments, which are linear combination of the corresponding CFF, were separated. The $\sin\phi$ moment for the neutron, which is sensitive to ${\cal{E}}_{n}$ was found very small, nevertheless comparison with models where different values  of the angular momentum of the $u$ and $d$ quarks, shows that the data are sensitive to this quantity as shown in Figure~\ref{fig:neutr}. An exploratory analysis of Hall B on ND$_{3}$ data is currently underway~\cite{daria}. 
Finally DVCS on nuclei was measured in Hall B. The experiment used a $^{4}{\rm He}$ target, which being a spin-0 nucleus  at twist-2 is sensitive only to the GPD $H_{A}$. The analysis, nearly completed, extracted BSA for both coherent ($e ^{4}{\rm He} \rightarrow e ^{4}{\rm He} \gamma$) and incoherent ($e ^{4}{\rm He} \rightarrow e p \gamma$) scattering. Preliminary results are shown in Figure~\ref{fig:nuclei}~\cite{hattawy}.  The bound proton shows a lower asymmetry relative to the free one in the different bins in $\Bx$.

The experimental program at Jefferson lab on DVCS has been producing numerous compelling results. Compton form factors for proton, neutron and helium were extracted in several kinematic bins offering a first insight on the distribution of the electric and axial charge for valence and sea quark and hopefully to extract the angular  momentum of the $d$ and $u$ quarks. To have a full picture of the GPDs, more data are needed. The newly upgraded Jefferson Lab accelerator at 12 GeV together with the upgraded detectors in the three halls will give a plethora of new data which will allow us to map three-dimensionally the nucleon.



\end{document}